\documentclass[aps,prl,twocolumn,floats,showpacs,superscriptaddress]{revtex4}
\usepackage{epsfig}

\begin{document}

\title{Dynamical and correlation properties  of the Internet}

\author{Romualdo Pastor-Satorras}
\affiliation{Departament  de F{\'\i}sica i Enginyeria Nuclear,
  Universitat Polit{\`e}cnica de Catalunya, Campus Nord B4,
  08034 Barcelona, Spain}
\author{Alexei V{\'a}zquez}
\affiliation{International School for Advanced Studies SISSA/ISAS, via
  Beirut 4, 34014 Trieste, Italy}
\author{Alessandro Vespignani}
\affiliation{The Abdus Salam International Centre for Theoretical Physics
  (ICTP), P.O. Box 586, 34100 Trieste, Italy}

\date{\today}

\begin{abstract}
  The description of the Internet topology is an important open
  problem, recently tackled with the introduction of scale-free
  networks.  In this paper we focus on the topological and dynamical
  properties of real Internet maps in a three years time interval. We
  study higher order correlation functions as well as the dynamics of
  several quantities.  We find that the Internet is characterized by
  non-trivial correlations among nodes and different dynamical
  regimes.  We point out the importance of node hierarchy and aging
  in the Internet structure and growth.  
  Our results provide hints towards the realistic modeling of the
  Internet evolution.
\end{abstract}

\pacs{PACS numbers: 89.75.-k,  87.23.Ge, 05.70.Ln}

\maketitle

Complex networks play an important role in the understanding of many
natural systems \cite{strog01,amaral}.  A network is a set of nodes and
links, representing individuals and the interactions among them,
respectively. Despite this simple definition, growing networks can
exhibits an high degree of complexity, due to the inherent wiring
entanglement occurring during their growth.  The Internet is a capital
example of growing network with technological and economical
relevance; however, the recollection of router-level maps of the
Internet has received the attention of the research community only
very recently \cite{nlanr,caida,lucent}.  The statistical analysis
performed so far have revealed that the Internet exhibits several
non-trivial topological properties.
(wiring redundancy, clustering, etc.).  
Among them, the presence of a power-law connectivity
distribution \cite{faloutsos,gcalda}  makes the Internet an
example of the recently identified class of scale-free networks
\cite{barab99}.

In this Letter, we focus on the dynamical properties of the Internet.
We shall consider the evolution of real Internet maps from 1997 to
2000, collected by the National Laboratory for Applied Network
Research (NLANR) \cite{nlanr}.
In particular, we
will inspect the correlation properties of nodes' connectivity, as
well as the time behavior of several quantities related to the growth
dynamics of new nodes.  Our analysis shows a dynamical behavior with
different growth regimes depending on the node's age and connectivity.
The analysis points out two distinct wiring processes; the first one
concerns newly added nodes, while the second is related to already
existing nodes increasing their interconnections.  A feature
introduced in the present work refers to the Internet hierarchical
structure, reflected in a non-trivial scale-free connectivity
correlation function.  Finally, we discuss recent models for the
generation of scale-free networks in the light of the present analysis
of real Internet maps.  
The results presented in this Letter could help developing more 
accurate models of the Internet.

Several Internet mapping projects are currently devoted to obtain
high-quality router-level maps of the Internet.  In most cases, the
map is constructed by using a hop-limited probe (such as the UNIX {\em
  trace-route} tool) from a single location in the network. In this
case the result is a ``directed'' map as seen from a specific point on
the Internet \cite{lucent}.  This approach does not correspond to a
complete map of the Internet because cross-links and other technical
problems (such as multiple IP aliases) are not considered. Heuristic
methods to take into account these problems have been
proposed~\cite{mercator}. However, it is not clear their reliability
and the corresponding completeness of maps constructed in this way. A
different representation of Internet is obtained by mapping the
autonomous systems (AS) topology. Each AS number approximately maps to
an Internet Service Provider (ISP) and their links are inter-ISP
connections. In this case it is possible to collect data from several
probing stations to obtain complete interconnectivity maps 
\cite{nlanr,caida}.
In particular, the NLANR project is collecting data since
Nov. 1997, and it provides topological as well as dynamical
information on a consistent subset of the Internet. The first Nov.
1997 map contains 3180 AS, and it has grown in time until the Dec.
1999 measurement, consisting of 6374 AS.  In the following we will
consider the graph whose nodes represent AS and whose links represent
the connections between AS.

In dealing with the Internet as an evolving network, it is important
to discern whether or not it has reached a stationary state whose
average properties are time-independent. 
As a first step, we have analyzed the behavior in time of several
average quantities such as the connectivity $\left<k\right>$, the
clustering coefficient $\left< C\right>$ and the average minimum path
distance $\left<d\right>$ of the network \cite{watts98}.  The first
two quantities (see Table I) show a very slow tendency to increase in
time, while the average minimum path distance is slowly decreasing
with time.  A more clear-cut characterization of the topological
properties of the network is given by the connectivity distribution,
$P(k)$.  In Fig. 1 we show the probability $P(k)$ that a given node
has $k$ links to other nodes.  We report the distribution for
snapshots of the Internet at different times. In all cases, we found a
clear power law behavior $P(k)\sim k^{-\gamma}$ with $\gamma=2.2\pm
0.1$. The distribution cut-off is fixed by the maximum connectivity
of the system and is related to the overall size of the Internet map.
On the other hand, the power
law exponent $\gamma$ seems to be independent of time and in good
agreement with previous measurements \cite{faloutsos}.  This evidence
seems to point out that the Internet's topological properties have
already settled to a rather well-defined stationary state.

\begin{table}[b]
\begin{ruledtabular}
\begin{tabular}{|c|c|c|c|}
Year & 1997 & 1998 & 1999\\
\hline
$\left\langle k\right\rangle$ & 3.47(4) & 3.62(5) & 3.82(6)\\
$\left\langle C\right\rangle$ & 0.18(1) & 0.21(2) & 0.24(1)\\ 
$\left\langle d\right\rangle$ & 3.77(1) & 3.76(2) & 3.72(1)
\end{tabular}

\caption{Average properties for three different years. $\left\langle
  k\right\rangle$ is the average connectivity. $\left\langle d
  \right\rangle$ is the minimum path distance $d_{ij}$ averaged over every
  pair of nodes $(i,j)$. $\left\langle C\right\rangle$ is the clustering
  coefficient $C_i$ averaged over all nodes $i$, where $C_i$ is defined as
  the ratio between the number of links between the  neighbors of $i$ and
  its maximum possible value $k_i(k_i-1)$. Figures in parenthesis indicate
  the statistical uncertainty from averaging the values of the
  corresponding months in each year.}

\label{tab:1} 
\end{ruledtabular}
\end{table}

Initially, the modeling of Internet considered algorithms based on its
static topological properties \cite{zeg}.  However, since the Internet
is the natural outcome of a complex growth process, the understanding
of the dynamical processes leading to its present structure must be
considered as a fundamental goal. From this perspective, the
Barab{\'a}si-Albert (BA) model, Ref. \cite{barab99,mendes99}, can be
considered as a major step forward in the understanding of evolving
networks.
Underlying the BA model is the preferential attachment rule
\cite{barab99}; i.e., new nodes will link
with higher probability to nodes with an already large connectivity.
This feature is
quantitatively accounted for by postulating that the probability of a
new link to attach to an old node with connectivity $k_i$, $\Pi(k_i)$,
is linearly proportional to $k_i$, $\Pi(k_i) \sim k_i$.
This is an intuitive feature of the Internet growth where large
provider hubs are more likely to establish connections than smaller
providers.  The BA model has been successively modified with the
introduction of several ingredients in order to account for
connectivity distribution with $2<\gamma<3$ \cite{barabba00,krap00}, local
geographical factors \cite{medina}, wiring among existing
nodes \cite{mendes00}, and age effects \cite{mendes01}.  While all these
models reproduce the scale-free behavior of the connectivity
distribution, it is interesting to inspect deeper the Internet's
topology to eventually find a few discriminating features of   
the dynamical processes at the basis of the Internet
growth.

\begin{figure}[t]
  \centerline{\epsfig{file=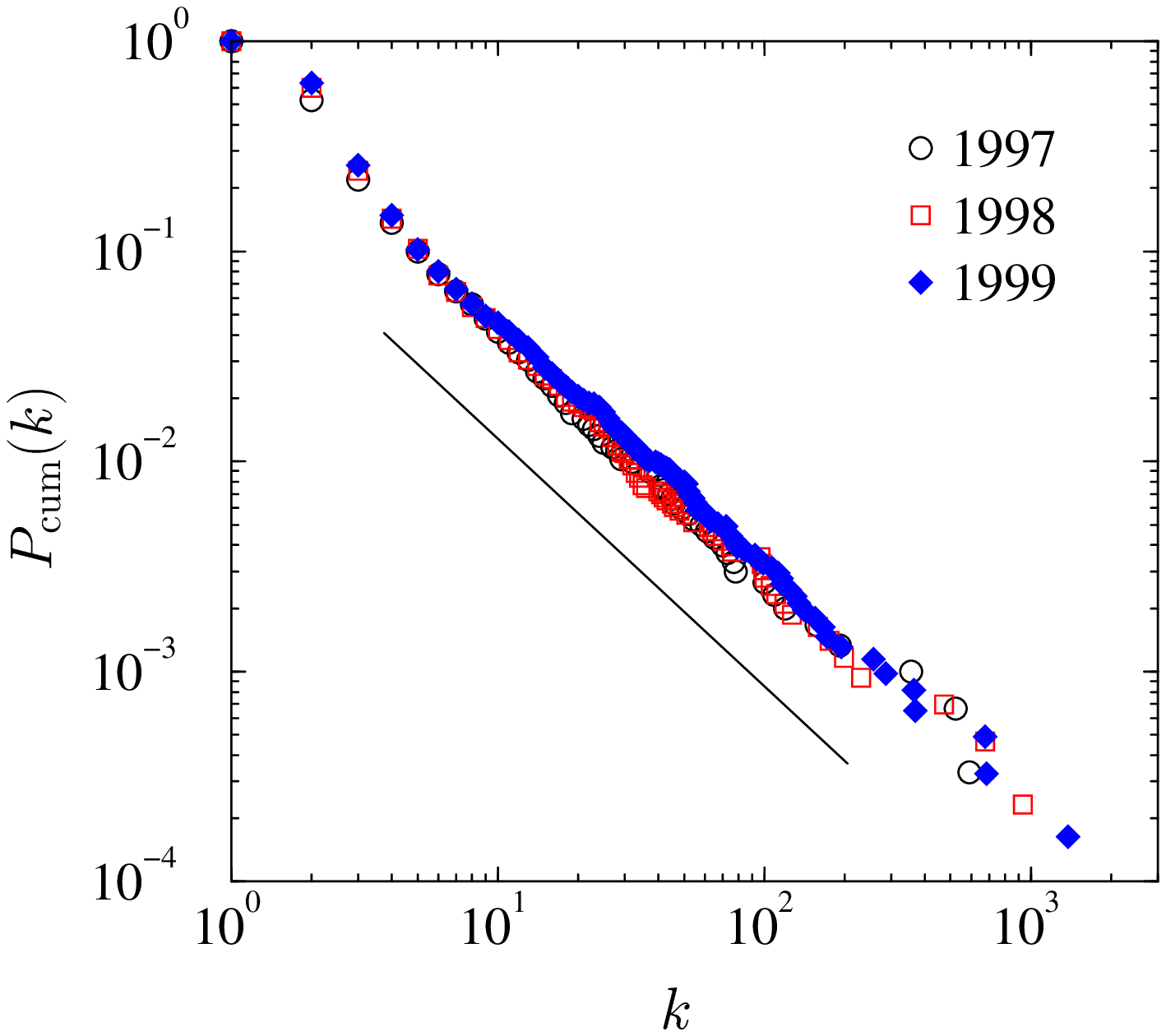, width=7cm}}
  \caption{Cumulated connectivity distribution 
    for the 1997, 1998, and 1999 snapshots of the Internet.  The power
    law behavior is characterized by a slope $-1.2$, which yields a
    connectivity exponent $\gamma=2.2$. }
\end{figure}

\begin{figure}[t]

  \centerline{\epsfig{file=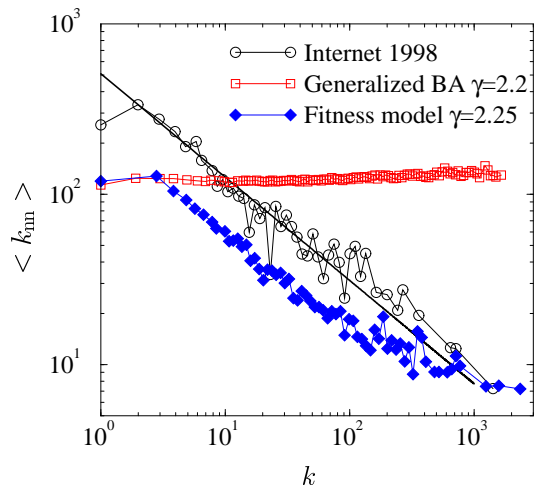, width=7cm}}
  
  \caption{Average connectivity $\left <k_{nn} \right>$ of the nearest
    neighbors of  
    a node depending on its connectivity $k$ for the 1998 snapshot of
    Internet, the generalized BA model with $\gamma=2.2$,
    Ref.~\protect\cite{barab99}, and 
    the fitness model, Ref.~\protect\cite{barabba01}. The full line
    has a slope $-0.5$. The scattered results for very large $k$ are due to 
    statistical fluctuations.}
\end{figure}

A first step in a more detailed characterization of the Internet
concerns the exploration of the connectivity correlations.  This
factor is best represented by the conditional probability $P_c(k'|k)$
that a link belonging to node with connectivity $k$ points to a node
with connectivity $k'$.  If this conditional probability is
independent of $k$, we are in presence of a topology without any
correlation among the nodes' connectivity. In this case, $P_c(k'|
k)=P_c(k')\sim k' P(k')$, in view of the fact that any link points to
nodes with a probability proportional to their connectivity.  On the
contrary, the explicit dependence on $k$ is a signature of non-trivial
correlations among the nodes' connectivity, and the possible presence
of a hierarchical structure in the network topology.  A direct
measurement of the $P_c(k'| k)$ function is a rather complex task due
to large statistical fluctuations. More clear indications can be
extracted by studying the quantity $\left<k_{nn}\right>= \sum_{k'}k'
P_c(k'| k)$; i.e.  the nearest neighbors average connectivity of nodes
with connectivity $k$. In Fig.  2, we show the results obtained for
the Internet map of 1998, that strikingly exhibit a clear power law
dependence on the connectivity degree $\left<k_{nn}\right>\sim k^{-\nu}$,
with $\nu\simeq 0.5$.  This result clearly implies the existence of 
non-trivial correlation properties for the Internet.  The primary
known structural difference between Internet nodes is the distinction
between {\em stub} and {\em transit} domains.  Nodes in stub domains
have links that go only through the domain itself. Stub domains, on
the other hand, are connected via a gateway node to transit domains
that, on the contrary, are fairly well interconnected via many paths.
In other
words, there is a hierarchy imposed on nodes that is very likely at
the basis of the above correlation properties.  As instructive
examples, we report in Fig. 2 the average nearest-neighbor
connectivity for the generalized BA model with $\gamma=2.2$ \cite{barabba00}
and the fitness model described in Ref.~\cite{barabba01}, with
$\gamma=2.25$, for networks with the same size than the Internet snapshot
considered. While in the first case we do not observe any noticeable
structure with respect to the connectivity $k$, in the latter we
obtain a power-law dependence similar to the experimental findings.
The general analytic study of connectivity correlations in growing
networks models can be found in Ref.~\cite{krap01}. 
A detailed discussion
of different models is beyond the scope of the present paper; however,
it is worth noticing that a $k$-structure in correlation functions, as
probed by the quantity $\left<k_{nn}\right>$, does not arise in all
growing network models. 

In order to inspect the Internet dynamics, we focus our attention on
the addition of new nodes and links into the maps. In the three-years
range considered, we have kept track of the number of links $\ell_{new}$
appearing between a newly introduced node and an already existing
node.  We have also monitored the rate of appearance of links
$\ell_{old}$ between already existing nodes.  In Table II we see that the
creation of new links is governed by these two processes at the same
time. Specifically, the largest contribution to the growth is given by
the appearance of links between already existing nodes.  This clearly
points out that the Internet growth is strongly driven by the need of
a redundancy wiring and an increased need of available bandwidth for
data transmission. 

\begin{figure}[t]
  \centerline{\epsfig{file=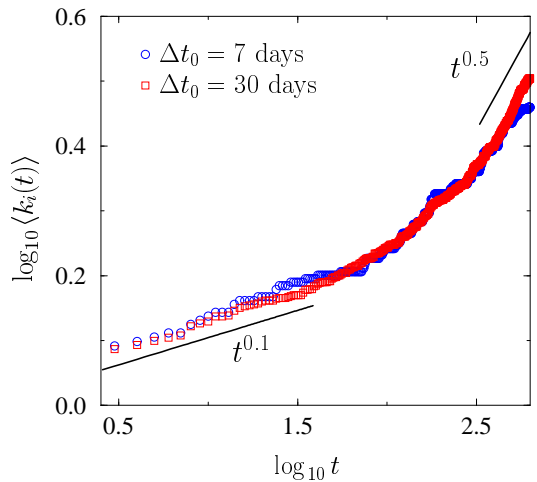, width=7cm}}
  \caption{Average connectivity of nodes borne within a small time
    window $\Delta t_0$, after a time $t$ elapsed since their appearance. 
    Time $t$ is measured in days.  As a comparison we report the lines
    corresponding to $t^{0.1}$ and $t^{0.5}$.}
\end{figure}

A customarily measured quantity in the case of growing networks is the
average connectivity $\left<k_i(t)\right>$ of new nodes as a function
of their age $t$. In Refs.~\cite{barab99,krap01} it is shown that
$\left<k_i(t)\right>$ is a scaling function of both $t$ and the
absolute time of birth of the node $t_0$. We thus consider the total
number of nodes born within an small observation window $\Delta t_0$, such
that $t_0\simeq {\rm const.}$ with respect to the absolute time scale that
is the Internet lifetime. For these nodes, we measure the average
connectivity as a function of the time $t$ elapsed since their birth.
The data for two different time windows are reported in Fig. 3, where
it is possible to distinguish two different dynamical regimes: At
early times, the connectivity is nearly constant with a very slow
increase ($\left<k_i(t)\right>\sim t^{0.1}$). Later on, the behavior
approaches a power law growth $\left<k_i(t)\right>\sim t^{0.5}$.  While
exponent estimates are affected by noise and limited time window
effects, the crossover between two distinct dynamical regimes is
compatible with the general aging form obtained in
Ref.~\cite{krap01}. In particular both the generalized BA model
\cite{barabba00} and the fitness model \cite{barabba01} 
present aging effects similar to those abtained 
in real data. A more detailed comparison would require a quantitative 
knowledge of the parameters to be used in the models and will
be reported elsewhere.

\begin{table}[b]
\begin{ruledtabular}
\begin{tabular}{|c|c|c|c|}
Year & 1997 & 1998 & 1999\\
\hline
$\ell_{new}$ & 183(9) & 170(8) & 231(11)\\
$\ell_{old}$ & 546(35) & 350(9) & 450(29)\\
$\ell_{new}/\ell_{old}$ & 0.34(2) & 0.48(2) & 0.53(3) 
\end{tabular}

\caption{Monthly rate of new links connecting existing nodes to  
new ($\ell_{new}$) and  old
($\ell_{old}$) nodes.}

\label{tab:2}
\end{ruledtabular}
\end{table}

A basic issue in the modeling of growing networks concerns
the preferential attachment hypothesis \cite{barab99}. Generalizing
the BA model algorithm it is possible to define models in which the
rate $\Pi(k)$ with which a node with $k$ links receive  new nodes is
proportional to $k^\alpha$. The inspection of the exact value of $\alpha$ in
real networks is an important issue since the connectivity properties
strongly depend on this exponent \cite{krap00,barabba01b}.  Here we use
a simple recipe that allows to extract the value of $\alpha$ by studying
the appearance of new links.  We focus on links emanating from newly
appeared nodes in different time windows ranging from one to three
years. We consider the frequency $\mu(k)$ of links that connect to nodes
with connectivity $k$.  By using the preferential attachment
hypothesis, this effective probability is $\mu(k)\sim k^\alpha P(k)$. 
Since we know that $P(k)\sim k^{-\gamma}$ we expect to find a power 
law behavior $\mu(k)\sim
k^{\alpha-\gamma}$ for the frequency. In Fig. 4, we report the obtained results
that shows a clear algebraic dependence $\mu(k)\sim k^{-1.2}$.  By using
the independently obtained value $\gamma=2.2$ we find a preferential
attachment exponent $\alpha\simeq 1.0$, in good agreement with the result
obtained with a different analysis in Ref.~\cite{barabba01b}.  We
performed a similar analysis also for links emanated by existing
nodes, recovering the same form of preferential 
attachment (see Fig. 4). 

\begin{figure}[t]

  \centerline{\epsfig{file=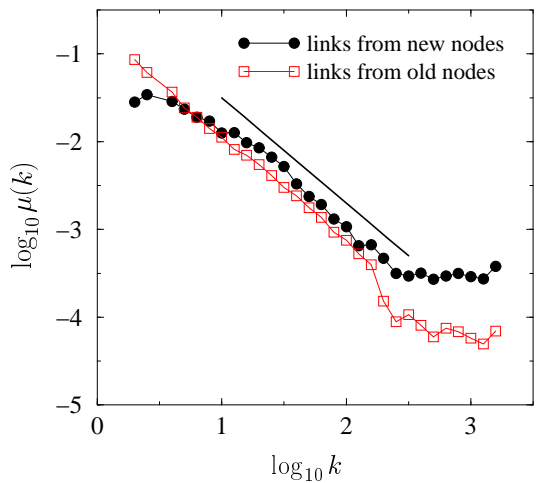, width=7cm}}
  \caption{Cumulative frequency of links emanating from new and 
        existing nodes that attach to nodes with connectivity $k$. 
        The straight line corresponds to a slope $-0.2$. The flat
        tail is originated from the poor statistics at the very high $k$ 
        values.  }
\end{figure}

In summary, we have shown that the Internet map exhibits a stationary
scale-free topology, characterized by non-trivial connectivity
correlations. An investigation of the Internet's dynamics confirms the
presence of a preferential attachment behaving linearly with the
nodes' connectivity and  identifies two different dynamical regimes
during the nodes' evolution. We point out that very likely
several other factors, such as the nodes' hierarchy, 
resource constraints and the
real geographical location of nodes can influence the Internet
evolution. The results reported here could be relevant 
for a more realistic modeling of the Internet growth and evolution and
could have implications in the study of the resilience to attacks and
spreading phenomena in this network \cite{barabbares,pv01}.

This work has been partially supported by the European Network 
Contract No. ERBFMRXCT980183.  RP-S also acknowledges support from the
grant CICYT PB97-0693. We thank  M.-C. Miguel, Y. Moreno-Vega, and  
R. V. Sol{\'e} for helpful comments and
discussions.


\begin{thebibliography}{10}

\bibitem{strog01}
        S.H Strogatz, Nature, {\bf 410 }, 268 (2001).
\bibitem{amaral} 
        L. A. N. Amaral, A. Scala, M. Barth\'{e}l\'{e}my, and
        H. E. Stanley, Proc. Nat. Acad. Sci. {\bf 97}, 11149 (2000).
\bibitem{nlanr}
        The National Laboratory for Applied Network Research (NLANR),
        sponsored by the National Science Foundation, provides 
        Internet routing related information based on BGP data
        (see http://moat.nlanr.net/).
\bibitem{caida}
        The Cooperative Association for Internet Data Analysis (CAIDA),
        located at the San Diego Supercomputer Center, 
        provides measurements of Internet traffic metrics 
        (see http://www.caida.org/home/).
\bibitem{lucent}
        B. Cheswick and H. Burch, Intenet mapping project at Lucent Bell 
        Labs (see http://www.cs.bell-labs.com/who/ches/map/).
\bibitem{faloutsos}
        M.~Faloutsos, P.~Faloutsos, and C.~Faloutsos, 
        ACM SIGCOMM '99, Comput.Commun. Rev. {\bf 29}, 251 (1999).
\bibitem{gcalda}
        G. Caldarelli, R. Marchetti, and L. Pietronero, 
        Europhys. Lett. {\bf 52}, 386 (2000).
\bibitem{barab99}
        A.-L. Barab\'{a}si and R.~Albert, Science
        {\bf 286}, 509 (1999);
        A.-L. Barab\'{a}si, R.~Albert, and H.~Jeong, 
        Physica A {\bf 272}, 173 (1999).
\bibitem{mercator}
        Mapping the Internet within the SCAN project at the 
        information Sciences Institute 
        (see  http://www.isi.edu/div7/scan/)
\bibitem{watts98}
D.~J. Watts and S.~H. Strogatz, Nature {\bf 393}, 440
(1998). 
\bibitem{zeg}
        E. W. Zegura, K. Calvert and M. J. Donahoo,
        EEE/ACM Transactions on Networking, December 1997. 
\bibitem{mendes99}
        S.~N. Dorogovtsev, J.~F. F.~Mendes, 
        and A.~N. Samukhin, Phys. Rev. Lett. {\bf 85}, 4633 (2000).
\bibitem{barabba00}
        R. Albert and A.-L. Barab\'{a}si,
        Phys. Rev. Lett. {\bf 85}, 5234 (2000).
\bibitem{krap00}
        P.L. Krapivsky, S. Redner and F. Leyvraz, 
        Phys. Rev. Lett. {\bf 85}, 4629 (2000).
\bibitem{medina}
        A.~Medina, I.~Matt, and J.~Byers, 
        Comput. Commun. Rev. {\bf 30}, 18 (2000).
\bibitem{mendes00}
        S.~N. Dorogovtsev and J.~F. F.~Mendes,
        Europhys. Lett. {\bf52}, 33 (2000).
\bibitem{mendes01}
        S.~N. Dorogovtsev and J.~F. F.~Mendes,
        Phys. Rev. E {\bf 63}, 025101(R) (2001).
\bibitem{barabba01}
        G.Bianconi and A.-L. Barab\'{a}si,
        cond-mat/0011029
\bibitem{krap01}
        P.L. Krapivsky and  S. Redner, 
        Phys. Rev. E {\bf 63}, 066123 (2001).
\bibitem{barabba01b}
        H. Jeong, Z. N\'{e}da and A.-L. Barab\'{a}si,
        cond-mat/0104131
\bibitem{barabbares}
        R.~A. Albert, H.~Jeong, and A.-L. Barab\'{a}si, Nature
        {\bf 406}, 378 (2000); D.~S. Callaway, M.~E.~J. Newman, 
        S.~H. Strogatz, and D.~J. Watts, Phys.Rev. Lett. 
        {\bf 85}, 5468 (2000); R.~Cohen, K.~Erez, 
        D.~{ben-Avraham}, and S.~Havlin, Phys. Rev. Lett.
        {\bf 86}, 3682 (2001).
\bibitem{pv01}
        R.~Pastor-Satorras and A.~Vespignani, 
        Phys. Rev. Lett. {\bf 86}, 3200 (2001).


\end{thebibliography}
\end{document}